# Origin of the monolayer Raman signature in hexagonal boron nitride: a first-principles analysis

*Jorge Ontaneda,[1] Anjali Singh,[2] Umesh V. Waghmare,[2] and Ricardo Grau-Crespo[1]\**

[1] Department of Chemistry, University of Reading, Whiteknights, Reading RG6 6AD, United Kingdom.

[2] Theoretical Sciences Unit, Jawaharlal Nehru Centre for Advanced Scientific Research (JNCASR), Bangalore 560064, India.

\* Email: r.grau-crespo@reading.ac.uk

**(13th March 2018)**

**ABSTRACT:** Monolayers of hexagonal boron nitride (*h*-BN) can in principle be identified by a Raman signature, consisting of an upshift in the frequency of the $E_{2g}$ vibrational mode with respect to the bulk value, but the origin of this shift (intrinsic or support-induced) is still debated. Herein we use density functional theory calculations to investigate whether there is an intrinsic Raman shift in the *h*-BN monolayer in comparison with the bulk. There is universal agreement among all tested functionals in predicting the magnitude of the frequency shift upon a variation in the in-plane cell parameter. It is clear that a small in-plane contraction can explain the Raman peak upshift from bulk to monolayer. However, we show that the larger in-plane parameter in the bulk (compared to the monolayer) results from non-local correlation effects, which cannot be accounted for by local functionals or those with empirical dispersion corrections. Using a non-local-correlation functional, we then investigate the effect of finite temperatures on the Raman signature. We demonstrate that bulk *h*-BN thermally expands in the direction perpendicular to the layers, while the intralayer distances slightly contract, in agreement with observed experimental behavior. Interestingly, the difference in in-plane cell parameter between bulk and monolayer decreases with temperature, and becomes very small at room temperature. We conclude that the different thermal expansion of bulk and monolayer partially "erases" the intrinsic Raman signature, accounting for its small magnitude in recent experiments on suspended samples.

## 1. INTRODUCTION

Hexagonal boron nitride (*h*-BN) is a layered material with structure similar to that of graphite, but with alternating B and N atoms, instead of C atoms, forming honeycomb-like networks.[1] Its properties have been attracting a great deal of interest over the last years[2] since the availability of *h*-BN samples with an atomically flat surface[3] facilitates studies of thin films applications based on *h*-BN. For instance, flakes of *h*-BN have been used as a thin dielectric to top-gate graphene in electronic devices,[4] as well as an inert substrate for graphene.[5] Song *et al*.[6] demonstrated that *h*-BN films consisting of two to five atomic layers can be synthetized via epitaxial growth on copper and then transferred onto a chosen substrate, opening up new opportunities to exploit its properties and potential applications in electronics, especially as an interesting two-dimensional defect-free dielectric material when sandwiched between conducting materials.[7]

The identification of the number of layers in mono-, bi- and few- layers of graphene is often done via either optical contrast[8] or Raman signatures.[9] The same two strategies have been discussed by Gorbachev and co-workers to identify mono- and few-layers *h*-BN.[10] However, because of its large bandgap (above 5 eV)[10–12] *h*-BN has zero opacity, i.e. it exhibits very weak optical contrast.[10] Raman spectroscopy seems to be a better route, particularly to avoid misidentification of mono- and bi-layers of *h*-BN as they look very similar under the microscope. Experimentally, bulk *h*-BN exhibits a characteristic Raman peak occurring at 1366 cm$^{-1}$ corresponding to an $E_{2g}$ phonon mode. Gorbachev *et al.*[10] observed that mono- and bilayers *h*-BN exhibited maximum upshifts of 4 cm$^{-1}$ and 1 cm$^{-1}$, respectively, which could serve as Raman signatures of these systems.

A previous theoretical study[13] based on density functional theory (DFT) calculations with the local density approximation (LDA), suggested that the difference between frequencies of Raman peaks in bulk *h*-BN and a single sheet is due to slightly elongated B-N bonds in the former, causing a softening of $E_{2g}$ phonon mode. However, as we will discuss in more detail below, the LDA functional employed in their work is unable to account correctly for the interlayer interactions in *h*-BN, in particular the dominant dispersion component, so a more sophisticated theoretical analysis is needed.

Recent experiments by Cai *et al.*[14] on suspended *h*-BN samples revealed much smaller upshifts (up to ~1 cm$^{-1}$) in the Raman frequency from bulk to monolayer. These authors suggested that the larger upshifts of ~4 cm$^{-1}$ observed in supported samples are actually due to the interaction with the support. According to this interpretation, the higher flexibility of the monolayer allows it to follow the uneven surface of the substrate more closely and gain more compressive strain; the intrinsic shift may be very small or non-existent.

Here we present a complete theoretical analysis, using DFT functionals that are able to describe correctly the dispersion interactions between layers, in an attempt to clarify whether there is an intrinsic upshift in the Raman frequency of the *h*-BN monolayer. We discuss the role of interlayer dispersion interactions in determining the relative cell parameters and Raman frequencies of bulk and monolayer, and calculate the effect of thermal expansion on the Raman signature of BN-monolayer.

## 2. COMPUTATIONAL METHODS

We performed density functional theory (DFT) calculations using planewave basis sets as implemented in VASP.[15,16] The projected augmented wave (PAW)[17,18] method was used to describe the interaction between the valence electrons and the core, keeping the 1*s* orbitals of both B and N frozen in the atomic reference configurations. As we will discuss below, the change in lattice parameter from bulk to monolayer is in the order of a fraction of picometer; therefore, we have used high precision parameters in our simulations to capture these subtle effects. In particular, we have employed the "hard" frozen-core potentials provided in VASP. Using the "normal" potential leads to the same trends described here, but introduces an error (in the order of 0.1 pm, which is small but significant for this study) in the relative values of the cell parameters of bulk vs. single-layer, compared to the results obtained with the hard potentials. Plane wave basis was truncated at a kinetic energy cutoff of 1050 eV (set at 50% above the default value for the potentials, in order to minimize Pulay errors).[19] In order to reduce the noise in the calculated forces, we use an additional support grid, with 8 times more points than the standard fine grid, for the evaluation of the augmentation charges. The maximum force on ions for geometry relaxations was set up to a very low threshold of $10^{-4}$ eV/Å (we checked that decreasing this value even more to $10^{-5}$ eV/Å, had no effect on the calculated geometries to the precisions reported here). Monkhorst-Pack grids[20] with a maximum separation of 0.35 Å$^{-1}$ between **k**-points were used in sampling the Brillouin-zone for integrations. This grid density corresponds to a 9×9×3 grid of **k**- points in the bulk *h*-BN. To keep interactions between periodic images small, a relatively large vacuum space of 15 Å was used between layers in the simulation of single-layer BN. Vibrational frequencies were obtained using a finite-differences method, where symmetry was employed to reduce the number of displacements.

We compared results obtained from calculations with various functionals including those based on the local density approximation (LDA)[21] and on the generalized gradient approximation (GGA) in the formulation by Perdew-Burke-Ernzerhof (PBE)[22], as well as their empirical corrections by Grimme's method (D2 and D3).[23,24] We also consider a set of functionals where dispersion is treated with explicit non-local correlation: the original vdW-DF method[25] (referred to as revPBE-vdW herein), the modified version vdW-DF2[26] (referred to as rPW86-vdW2 herein) and three of the "opt" series (optB88-vdW, optB86b-vdW and optPBE-vdW) where the exchange functionals were optimized for the correlation part,[27] as developed and implemented in VASP by Klimeš *et al.*[28] Interlayer binding energies $E_b$ for bulk *h*-BN were calculated as the energy difference (per atom) between bound and separated layers:

$$E_\mathrm{b} = \frac{E_\mathrm{bulk} - 2E_\mathrm{ML}}{4} \quad (1)$$

where $E_\mathrm{ML}$ and $E_\mathrm{bulk}$ and are the energies of the monolayer cell (containing 2 atoms) and of the bulk cell (containing 2 layers and 4 atoms).

We have calculated the equilibrium structure of *h*-BN at finite temperatures by minimizing the vibrational free energy:

$F(\{a_i\}, T) = E(\{a_i\}) + F_\mathrm{vib}\bigl(\omega_{\mathbf{q},j}(\{a_i\}), T\bigr)$

$$= E(\{a_i\}) + \sum_{\mathbf{q},j} \frac{hc\omega_{\mathbf{q},j}(\{a_i\})}{2} + k_B T \sum_{\mathbf{q},j} \ln\left[1 - \exp\left(-\frac{hc\omega_{q,j}(\{a_i\})}{k_B T}\right)\right] \quad (2)$$

with respect to the cell parameters $\{a_i\}$. In this equation, $E(\{a_i\})$ corresponds to the ground-state energy of the structure at a given set of lattice parameters, $\omega_{\mathbf{q},j}(\{a_i\})$ is the frequency in cm$^{-1}$ of the $j^{th}$ phonon band at the point $\mathbf{q}$ in the Brillouin zone, $h$ is Planck's constant, $c$ is the speed of light, k$_B$ is Boltzmann's constant, and $T$ is the absolute temperature. This is called the quasiharmonic approximation (QHA), because the dependence of the phonon frequencies on the structural parameters introduces anharmonic effects in the calculation and permits the prediction of thermal expansion. Since the frequency of each mode in the Brillouin zone varies linearly with the cell parameters to a good approximation within the region of interest, we performed phonon calculations at a small grid of cell parameters and fit linear equations for all modes, as previously done by Mounet and Marzari in the simulation of the thermal behavior of graphene and graphite.[29] Combined with a polynomial expansion of the energies around the equilibrium point, this procedure allows us to define an analytical expression for the free energy, which can be then minimized at any given temperature.

3. **RESULTS AND DISCUSSION**

### 3.1. Variation of Raman frequency with in-plane cell parameter

The E$_{2g}$ is a doubly-degenerate Raman-active mode which involves the in-plane vibration of B and N atoms in opposite directions, as represented in the inset of **Figure 1**. As a starting point, we have calculated the E$_{2g}$ Raman frequencies of a single layer as a function of cell parameter, with each of the functionals. The results are summarized in **Figure 1**. The absolute value of the frequency predicted at a given cell parameter (say at the experimental room temperature cell parameter of the bulk $a$=2.5047 Å),[30] varies with the functional. However, results from all functionals are quite similar in the magnitude of the rate of frequency change with lattice parameter, which is approximately 22 cm$^{-1}$/pm. The Raman frequencies increase with the decrease in cell parameter, which is expected since a lattice contraction means shortening and stiffening of the B–N bond, producing a hardening of the $E_{2g}$ mode.

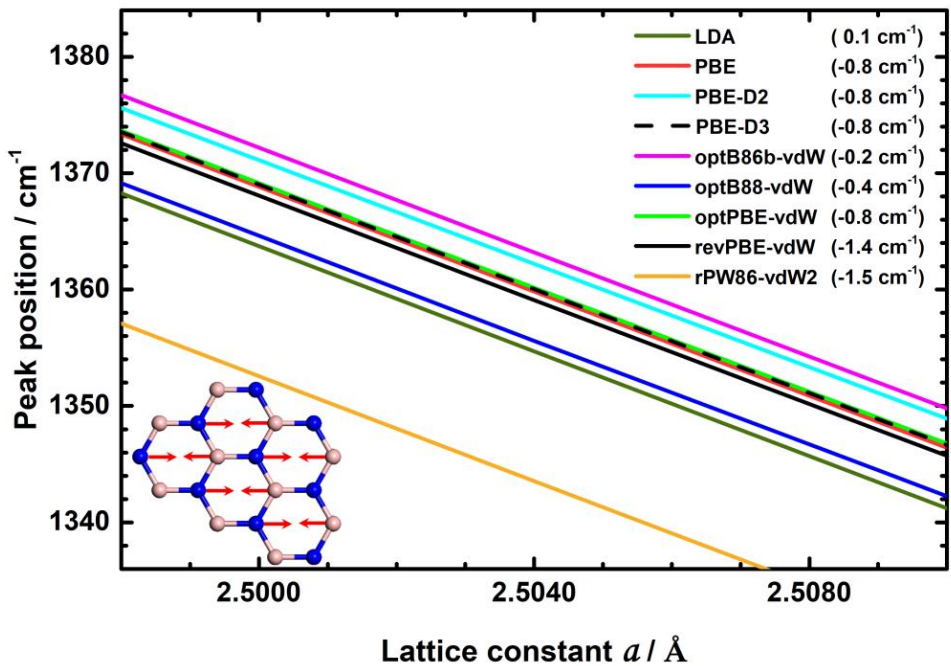

**Figure 1.** Relationship between Raman frequencies and cell parameters of single-layer boron nitride, as calculated with different density functionals. The behavior of the bulk is identical, except for small constant shifts (given in brackets in the legend).

We now present results of analogous calculations for the bulk, keeping the $c$ parameter initially fixed (we used here the experimental value extrapolated to zero temperature, $c$=6.60002 Å, in Ref. 30; the results of full geometry relaxations will be discussed next). For any of the functionals used here, there is very little variation of frequency between the bulk and the single layer at a given cell parameter. In fact, the frequency vs. cell parameter lines calculated for the bulk cannot be distinguished in a plot from those calculated for the single layer, and we have not plotted them separately. There is only a small shift from single layer to bulk (values given in brackets in **Figure 1**), which is negative for all functionals (except for LDA), and below 1 cm$^{-1}$ for most functionals (except the original vdW-DF functionals, for which the shifts are only slightly larger). This result, which is not surprising as E$_{2g}$ is an in-plane mode, makes it clear that the experimental frequency difference between single layer and bulk cannot be attributed *directly* to the interlayer interactions: it must be the result of the difference in cell parameter between the two systems.

The frequency of E$_{2g}$ mode of the bulk calculated at the experimental cell parameter is slightly below its experimental value for the bulk (1366 cm$^{-1}$)[10] for all functionals (only the rPW86-vdW2 exhibits a relative large deviation of 24 cm$^{-1}$). However, we are not interested here in predicting the absolute values (which in any case are very sensitive to cell parameter variations due to thermal effects not included in our calculations), but on understanding the relative values between the bulk and the few-layer systems.

### 3.2. Variation of equilibrium lattice parameter from monolayer to bulk

The analysis above suggests that the experimentally observed upshifts in the Raman frequency from bulk to monolayer could be due to a small contraction of the lattice parameter of the latter with respect to the former, which would be in agreement with the conclusion reached in references 10 and 13. However, the contraction required to explain the even the largest experimental signature reported for the monolayer (4 cm$^{-1}$) is quite small, of 0.2 pm. We will now consider whether DFT simulations reproduce this small lattice contraction based on equilibrium geometry calculations via energy minimizations.

The comparison between the equilibrium geometries of bulk and single layer obviously requires a correct description of the bulk interlayer attraction, which, like in graphite, has two main components: 1) the dispersion interaction, and 2) the electronic kinetic energy reduction due to an increased delocalization of the $2p_z$ orbitals between adjacent layers.[31] In the case of *h*-BN the atoms are slightly charged due to the polarity of the B-N bond, so there is also a small contribution from direct Coulombic interactions between layers. It is well known for graphite that GGA functionals like PBE cannot account for the interlayer attraction, whereas the LDA does give a reasonable interlayer potential with a minimum close to the experimental value.[31,32] **Figure 2** shows the comparison between the LDA, PBE and optB88-vdW interlayer potentials at constant lateral cell parameter for *h*-BN. As for graphite, the LDA interlayer potential for *h*-BN exhibits a well-defined minimum close to the experimental value of the interlayer distance. This "success" of the LDA masks its actual inability to properly describe the physics of the interlayer interaction. In fact, neither the GGA nor the LDA are able to account for long-range dispersion interactions, which are non-local correlation effects. In the LDA, however, the kinetic energy reduction effect is exaggerated by the tendency of this functional to overdelocalize the charge density. Because of the cancellation of these errors, the LDA interlayer potential mimics the correct one, albeit with a smaller binding energy and a too fast falloff at long distances, compared to functionals including non-local correlations.

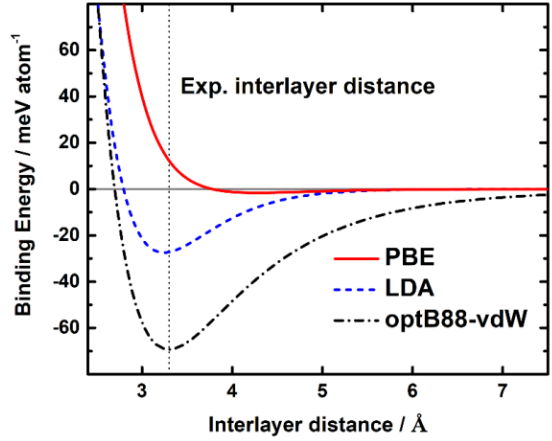

**Figure 2.** Interlayer binding energy (per atom) of bulk *h*-BN as a function of interlayer distance (*c*/2) calculated by LDA, PBE and optB88-vdW. In all these calculations, the lattice constant *a* was fixed to the bulk zero-Kelvin experimental value (2.5059 Å).

**Table 1.** Calculated equilibrium lattice parameters (ignoring vibrational effects), Raman-active ($E_{2g}$) frequencies, and interlayer binding energies of bulk *h*-BN using different functionals, and comparison with room-temperature experimental values.

| Functional | *a* (Å) | *c* (Å) | Deviation from experiment | | $\omega[E_{2g}]$ (cm$^{-1}$) | $E_b$ (meV/atom) |
|---|---|---|---|---|---|---|
| | | | $\Delta a/a_{exp}$ (%) | $\Delta c/c_{exp}$ (%) | | |
| Experiment | 2.5059[30] | 6.6000[30] | - | - | 1366[10] | |
| LDA | 2.4911 | 6.4886 | -0.6 | -1.7 | 1383.8 | -27.1 |
| PBE | 2.5119 | 8.5022 | 0.2 | 28.8 | 1342.2 | -1.6 |
| PBE-D2 | 2.5082 | 6.1716 | 0.1 | -6.5 | 1350.5 | -76.9 |
| PBE-D3 | 2.5086 | 6.7793 | 0.1 | 2.7 | 1349.0 | -49.0 |
| optB86b-vdW | 2.5120 | 6.5170 | 0.2 | -1.3 | 1344.8 | -69.8 |
| optB88-vdW | 2.5108 | 6.5896 | 0.2 | -0.2 | 1339.8 | -69.5 |
| optPBE-vdW | 2.5168 | 6.7884 | 0.4 | 2.9 | 1331.0 | -63.4 |
| revPBE-vdW | 2.5235 | 7.1011 | 0.7 | 7.6 | 1315.2 | -52.4 |
| rPW86-vdW2 | 2.5217 | 6.9814 | 0.6 | 5.8 | 1303.6 | -50.4 |

**Table 1** summarizes the relaxed cell parameters of bulk $h$-BN as obtained from several functionals considered here. In all cases, the $a$ lattice parameter is reasonably well described. However, the performances of these functionals in predicting the $c$ lattice parameter are quite different. In the case of GGA-PBE, the very weak interlayer interaction leads to too large a value of $c$. The empirically-corrected versions show an interesting behavior: PBE-D2 underestimates the interlayer distance by 7.3% whereas its redefined version, PBE-D3, overestimates it by only 1.8% which marks a significant improvement in the description of interlayer interactions. Among the non-local-correlation (NLC) functionals, the optB88-vdW gives the most accurate value of $c$ lattice parameter, which corresponds to a deviation of -1.1%.

We also report in **Table 1** the interlayer binding energies calculated with each functional at the corresponding equilibrium geometries. There are no experimental values for the binding energy of $h$-BN. Our calculations are in good agreement with those previously reported by Graziano *et al.* using a similar list of functionals.[33] Based on the comparison with graphite calculations, these authors concluded that the interlayer binding energy from non-local-correlation functionals were the most sensible, whereas other empirically corrected functionals, like PBE-D2 and also the Tkatchenko-Scheffler-corrected PBE, slightly overestimated the binding energies.

We now analyze the geometry of an isolated layer after relaxation, in comparison with the bulk. **Table 2** lists the $a$ lattice parameters obtained from calculations with each functional, and the relative values with respect to the bulk. All the NLC functionals (and the LDA) predict small contractions (between 0.1 and 0.2 pm) with respect to the bulk, which is the expected trend if the experimental Raman signature is interpreted as resulting from a lattice contraction. Interestingly, the Grimme functionals PBE-D2 and PBE-D3 predict instead a small expansion with respect to bulk values. The GGA-PBE frequencies are almost identical for bulk and single layer, since the optimization of the bulk geometry with this functional leads to a very large (unrealistic) interlayer separation. **Table 2** also shows the absolute values of the Raman frequencies calculated at the equilibrium geometries in both bulk and single-layer $h$-BN. The small contractions from bulk to single layer predicted by the non-local-correlation functionals can in principle account for the experimental upshift in Raman frequency. The magnitude of the variations in Raman frequencies predicted by the "opt" series of functionals, from 3.7 to 4.5 cm$^{-1}$, is in agreement with the values obtained from experiments in supported samples, although significantly higher than the experimental upshifts in suspended samples (up to ~1 cm$^{-1}$).[14] In contrast, PBE-D2 and PBE-D3 predict a downshift in frequency from bulk to single layer, in disagreement will all experiments.

**Table 2.** Calculated equilibrium lattice parameter (ignoring vibrational effects) and Raman active ($E_{2g}$) frequencies of single-layer $h$-BN using different functionals, and comparison with corresponding values for the bulk.

| Functional | Lattice parameter | | $\omega[E_{2g}]$ (cm$^{-1}$) | |
|---|---|---|---|---|
| | $a$ (Å) | Change from bulk (pm) | 1L | Change from bulk |
| **Experiment**[10] | - | - | 1370 | 4 |
| **LDA** | 2.4897 | -0.14 | 1387.1 | 3.3 |
| **PBE** | 2.5119 | -0.01 | 1342.3 | 0.1 |
| **PBE-D2** | 2.5121 | 0.39 | 1344.2 | -6.4 |
| **PBE-D3** | 2.5108 | 0.22 | 1344.8 | -4.2 |
| **optB86b-vdW** | 2.5102 | -0.18 | 1349.3 | 4.5 |
| **optB88-vdW** | 2.5091 | -0.17 | 1344.2 | 4.4 |
| **optPBE-vdW** | 2.5154 | -0.14 | 1334.7 | 3.7 |
| **revPBE-vdW** | 2.5225 | -0.10 | 1318.1 | 2.9 |
| **rPW86-vdW2** | 2.5207 | -0.10 | 1306.4 | 2.8 |

**3.3. Role of non-local correlation in the lattice expansion from monolayer to bulk**

In order to understand the different prediction of the monolayer-to-bulk cell variation from Grimme-corrected and NLC functionals, we have analyzed the variation in the energy of the bulk, calculated with either the PBE-D3 or the optPB88-vdW functional, upon contraction/expansion of the layer plane, at a fixed $c$ parameter. It is possible to decompose that energy variation into monolayer contributions (i.e. twice the energy variation obtained in the calculation of the isolated layer) and interlayer interaction effects (i.e. contributions to the binding energy $E_b$). An energy decomposition analysis (see Supplementary Information for details) reveals that, in the PBE-D3 calculations, the dispersion component (given by the Grimme term) is the dominant contribution to the inter-layer energy variation (the magnitude of the contributions from the other, non-dispersive interactions is only ~10% of the dominant contribution). On the other hand, in the optB88-vdW calculations, the dominant contribution to the inter-layer energy variation is the non-local correlation term (in this case the other contributions represent ~30% of the dominant one).

**Figure 3** shows the energy variation with $a$ for bulk and monolayer, as well as the contribution of the dispersion or NLC terms to the interlayer interactions. The difference in behavior between the inter-layer dispersion interaction term in the Grimme-corrected functional and the inter-layer non-local correlation term in the NLC functionals can explain the different output from the two types of functional.

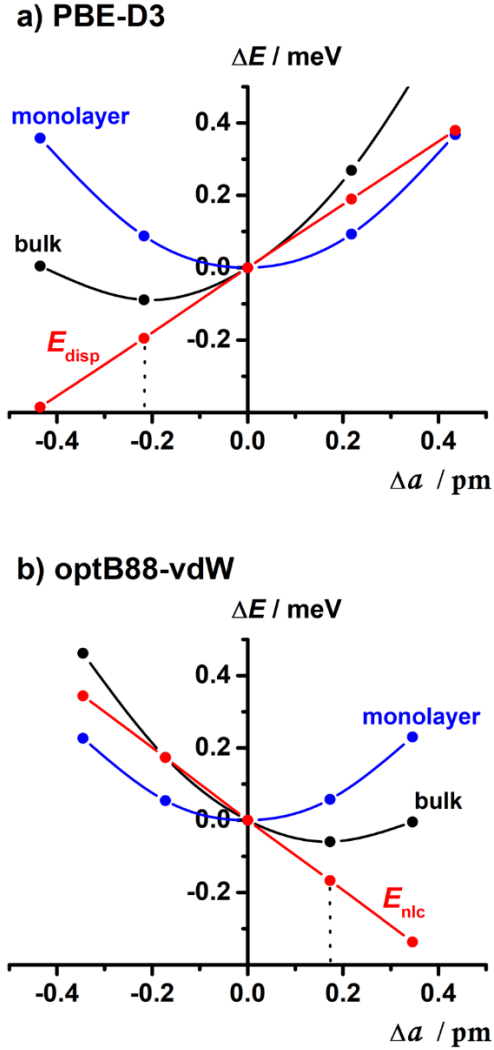

**Figure 3.** Variation of the monolayer and bulk energies around the equilibrium in-plane lattice parameter. The reference for the energies are the values of the given energy component at the cell parameter at which the monolayer is in equilibrium (for each functional).

In the case of PBE-D3, one can expect that the inter-layer dispersion interaction, given the simple mathematical definition of the Grimme correction as an attractive force, will tend to contract the layers, and this is indeed our observation. However, in the optB88-vdW, the non-local correlation does not behave in the same way, and it tends to expand the layers. The fact that the description of the in-plane expansion in the bulk requires a proper treatment of non-local electron correlation is interesting. Our analysis indicates that this expansion cannot be accounted for by invoking a simple picture of intra-plane bond weakening due to the interaction of $p_z$ electrons. The effect is more complex and requires the consideration of non-local (inter-plane) correlation effects.

Considering that the calculation of the relevant interactions is more sophisticated and accurate in the NLC functionals than in the Grimme-corrected GGA ones, we believe that the small lattice expansion of the bulk (in comparison with the monolayer) predicted by the former type of functional is reliable, at least in the low-temperature limit (the effect of finite temperatures will be discussed in section 3.5). The LDA is able to mimic the correct behavior, but based on the wrong physics as explained above.

### 3.4. Variation of Raman frequency with the number of layers

Using the optB88-vdW functional, we now consider the behavior of *N*-layer *h*-BN structures, with *N* up to 5. **Figure 4** shows the calculated frequency shifts with respect to the bulk, in comparison with the experimental data by Cai et al.[14] on both supported and suspended samples. In bilayers, we predict an upshift of 2.4 cm$^{-1}$ in the Raman frequency because of the contraction in the *a* parameter by 0.10 pm with respect to bulk. That value is in the middle of the range measured by Cai *et al.* on supported samples (between 1.8 and 3.2 cm$^{-1}$), but significantly above the values measured for suspended samples (~0.5 cm$^{-1}$). In an earlier experiment, Gorbachev *et al.* had measured a range of downshifts and upshifts for supported bilayers with respect to bulk,[10] but interpreted the maximum upshift of ~1 cm$^{-1}$ as the intrinsic value for the bilayer, with the other values arising from downshifts due to the interaction with the support. The calculated upshift for the trilayer 1.4 cm$^{-1}$ (contraction of 0.06 pm with respect to bulk) is close to the maximum value obtained for supported samples in Ref. 14, but again above the values measured in suspended samples. In the case of 4L and 5L, we have found upshifts of 1.1 cm$^{-1}$ (0.05 pm) and 1.0 cm$^{-1}$ (0.04 pm), respectively, i.e. the differences with respect to the bulk decrease with the increase in the number of layers $N$, as expected. The relative peak position $\Delta\omega$ is found to be proportional to $1/N$ (correlation $R^2$=0.997), with a proportionality constant of 4.5 cm$^{-1}$ (**Figure 4 inset**).

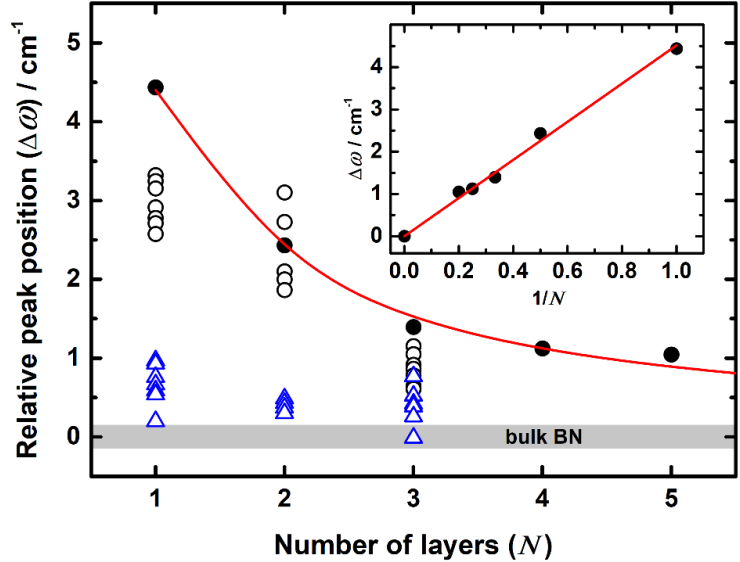

**Figure 4.** Relative positions of the Raman peaks for few-layer $h$-BN, with respect to the bulk, as obtained from optB88-vdW calculations ignoring vibrational contributions (solid circles). The red line is the fitting of a $1/N$ law (see inset); the open circles and open blue triangles are the experimental values reported by Cai *et al.* in Ref. 14 for supported and suspended samples, respectively.

### 3.5. Vibrational effects: finite-temperature calculations

All the results presented above were obtained ignoring vibrational contributions and therefore any thermal expansion effects. However, due to the small magnitude of the changes in cell parameters between bulk and monolayer involved in the present discussion, thermal effects are likely to be important. To account for these, we have performed temperature-dependent optimizations of the bulk and monolayer cell parameters, using the QHA.

The results are shown in **Figure 5**. Similarly to what was reported for graphite in Ref. 29, in *h*-BN the out-of-plane parameter (*c*) expands significantly with temperature (~5.60 pm from zero to room temperature), while the in-plane parameter *a* contracts, but only a little (~0.16 pm from zero to room temperature). Our results agree very well with the experimentally measured variation of bulk *h*-BN cell parameters with temperatures between 0 and 300 K as reported by Paszkowicz *et al.* in Ref. 30 (discounting the discrepancies in absolute values). The in-plane lattice contraction of bulk *h*-BN with temperature was first reported by Pease in Ref. 34. The physics behind this effect (in bulk and monolayer *h*-BN) can be expected to be the same as in graphite/graphene, i.e. to be caused primarily by low-frequency flexural phonons (acoustic out-of-plane modes).[35] However, the contraction of monolayer *h*-BN between 0 and 300 K is less than half of that in graphene over the same temperature range, which can be explained by the more negative Grüneisen parameters of the flexural modes of the latter[29] compared to the former.[14]

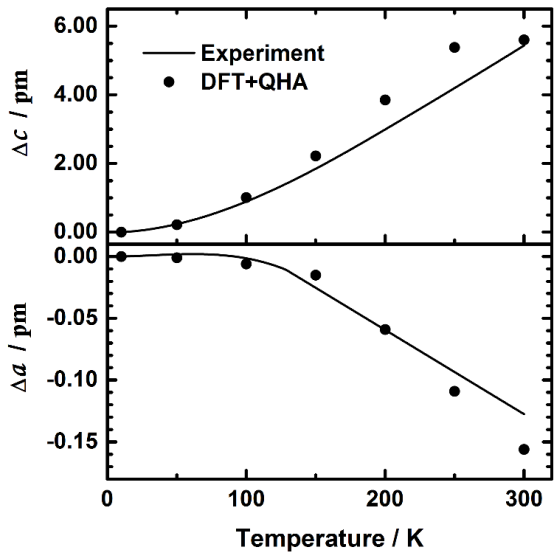

**Figure 5.** Relative dependence of bulk *h*-BN lattice parameters on temperature. The solid lines represent polynomial fittings to the experimental data as given in Ref. 30.

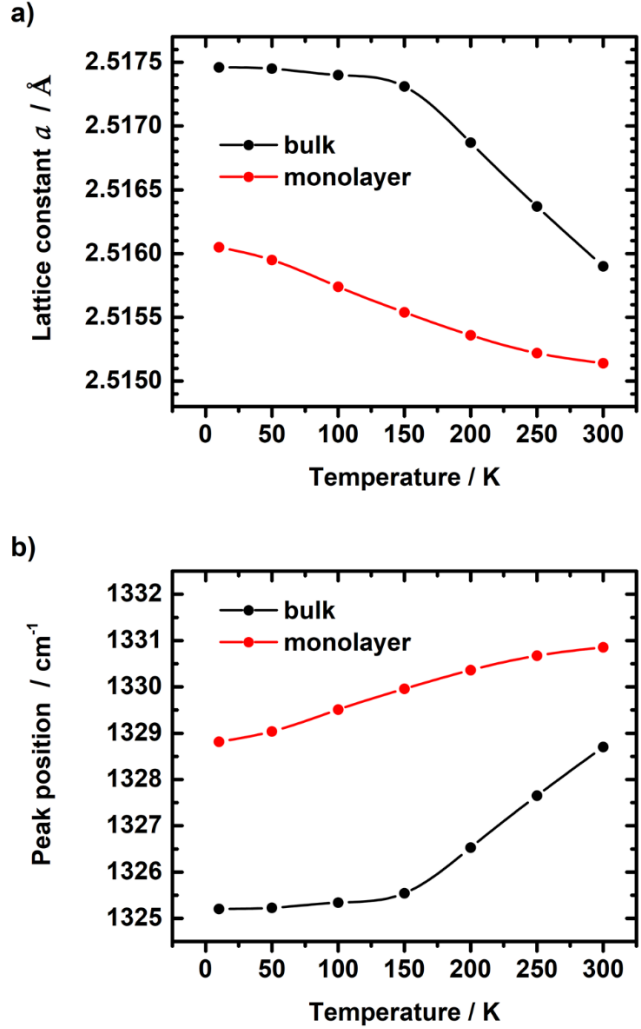

**Figure 6.** Predicted temperature variation of a) the in-plane lattice parameter and b) the Raman peak position of both bulk and monolayer *h*-BN, from DFT+QHA calculations.

There is a noticeable difference between the variation of the cell parameter with temperature for the bulk and the monolayer of *h*-BN. The thermal contraction in the monolayer is smaller, less than 0.1 pm from zero to 300 K. In the bulk, most of the in-plane contraction from zero to 300 K happens above 100 K, when the negative rate of variation in cell parameter with temperature increases. This effect can be interpreted as resulting from the expansion of the *c* parameter in the bulk: as *c* expands, the atomic layers in the bulk become more like the isolated layers (in terms of electrostatic/vdW environment), so the pace of in-plane contraction in the bulk has to increase to tend towards the *a* parameter value in the isolated layer.

This observation is important to our understanding of the experimentally observed Raman signatures of the BN monolayer. It is clear that at room temperature, a significant part of the difference in *a* cell parameter (and therefore the difference in Raman frequency, as shown in **figure 6b**) between bulk and single layer disappears as the results of thermal effects. This could be a possible explanation for the small difference in Raman frequency between bulk and monolayer (i.e. weak monolayer Raman signature) in the experiment with suspended samples by Cai *et al.*[14] Although our prediction for the intrinsic Raman signature of the monolayer at 300 K (~2 cm$^{-1}$) is still higher than the values of up to 1cm$^{-1}$ measured in suspended samples, our calculations including thermal effects do tend to support the conclusion by these authors of a weak intrinsic Raman signature. This interpretation also implies that the stronger Raman signature (4 cm$^{-1}$) observed previously in supported samples is not intrinsic and is probably due to support effects. Thus, the agreement between our calculations without thermal expansion and the experimental measurements on the supported samples (shown in **Figure 4**) is likely to be fortuitous: the absence of finite-temperature effects in that model made the Raman signature stronger, bringing it into agreement with the support-induced value.

## 4. CONCLUSIONS

In summary, we have gained new insights into the origin of the Raman signature of *h*-BN monolayer (difference in Raman frequency with respect to the bulk). First, we demonstrate that the origin of the Raman signature is clearly related to a difference in in-plane cell parameter between the bulk and the monolayer. We prefer this statement over the conclusion (e.g., in Ref.

14) that the Raman frequencies are determined by the level of strain. The bulk and the single layer have slightly different equilibrium cell parameters, which means that both systems will have different Raman frequencies in the absence of strain. Our calculations show that the "direct" effect of inter-layer interactions on Raman frequency (i.e. the effect arising without considering the difference in cell parameter) is very small; most of the effect arises from the lattice contraction.

We have shown that the best dispersion-corrected functionals (those with explicit non-local correlation) indeed predict, in a zero-temperature calculation, a contraction of lattice parameter from bulk to monolayer that can quantitatively explain the Raman shift observed in supported samples. While this could be at first interpreted as a confirmation of the existence of an intrinsic Raman shift, the simulation of the behavior at finite temperatures add a caveat: we show that the different thermal expansion of bulk and monolayer partially "erases" the intrinsic Raman signature. Our final conclusion of a weak intrinsic Raman signature is in agreement with the most recent experimental results using suspended samples by Cai *et al.*[14], and tend to support the conclusion by these authors that the stronger Raman signature in supported samples is predominantly a substrate-induced effect.

**ACKNOWLEDGMENTS**

J.O. acknowledges funding from Ecuador Government's agency SENESCYT in the form of a PhD studentship award (CA-2012-2). The authors thank the UK-India Education and Research Initiative (UKIERI) for funding this collaborative project through a Trilateral Partnership Grant IND/CONT/2013-14/054. This work made use of ARCHER, the UK's national high-performance computing service, *via* the UK's HPC Materials Chemistry Consortium, which is funded by EPSRC (EP/L000202).

# SUPPLEMENTARY INFORMATION

We provide here the details of the variation of the interlayer energy in the *h*-BN bulk, calculated with either the PBE-D3 or the optPB88-vdW functional, upon contraction/expansion of the layer plane, at a fixed *c* parameter. The purpose is to demonstrate that the most important contribution to variations in the interlayer energy comes from dispersion interactions.

As described in the article, the variation of bulk energy with *a* can be decomposed into monolayer contributions (i.e. twice the energy variation obtained in the calculation of the isolated layer) and interlayer interaction effects (i.e. contributions to the binding energy). The interlayer energy can in turn be decomposed into dispersion contributions (from Grimme's term in PBE-D3 or from non-linear correlation term in optB88-vdW) and other contributions. **Table 1S** shows that, in the PBE-D3 calculations, the dispersion component (given by the Grimme term) is the dominant contribution to the inter-layer energy variation (the magnitude of the contributions from the other, non-dispersive interactions, is only ~10% of the dominant contribution). Analogously, in the optB88-vdW calculations, the dominant contribution to the inter-layer energy variation is the non-local correlation term (in this case the other contributions represent ~30% of the dominant one).

**Table 1S.** Variation of the interlayer energy ($\Delta E_{\text{inter}}$) in bulk $h$-BN with in-plane lattice parameter as calculated with the **PBE-D3** functional. $\Delta E_{\text{inter}}$ is decomposed into two terms: $\Delta E_{\text{disp}}$ is the dispersion contribution and $\Delta E_{\text{rest}}$ contains the rest of the contributions. Here $a_0 = 2.5086$ Å is the equilibrium cell parameter of single-layer $h$-BN with this functional, and $|\Delta a|=0.0022$ Å is the absolute value of the difference in equilibrium cell parameter between bulk and monolayer. The interlayer energies per formula unit are given relative to their values at cell parameter $a_0$.

| Cell parameter | $\Delta E_{\text{inter}}$ (meV) | $\Delta E_{\text{disp}}$ (meV) | $\Delta E_{\text{rest}}$ (meV) |
|---|---|---|---|
| $a_0 - 2|\Delta a|$ | -0.35 | -0.38 | 0.03 |
| $a_0 - |\Delta a|$ | -0.18 | -0.19 | 0.02 |
| $a_0$ | 0.00 | 0.00 | 0.00 |
| $a_0 + |\Delta a|$ | 0.18 | 0.19 | -0.01 |
| $a_0 + |\Delta a|$ | 0.35 | 0.38 | -0.03 |

**Table 2S.** Variation of the interlayer energy ($\Delta E_{\text{inter}}$) in bulk $h$-BN with in-plane lattice parameter as calculated with the **PBE-D3** functional. $\Delta E_{\text{inter}}$ is decomposed into two terms: $\Delta E_{\text{nlc}}$ is the non-linear correlation contribution and $\Delta E_{\text{rest}}$ contains the rest of the contributions. Here $a_0 = 2.5108$ Å is the equilibrium cell parameter of single-layer $h$-BN with this functional, and $|\Delta a|=0.0017$ Å is the absolute value of the difference in equilibrium cell parameter between bulk and monolayer. The bulk energies per formula unit are given relative to their values at cell parameter $a_0$.

| Cell parameter | $\Delta E_{\text{inter}}$ (meV) | $\Delta E_{\text{nlc}}$ (meV) | $\Delta E_{\text{rest}}$ (meV) |
|---|---|---|---|
| $a_0 - 2|\Delta a|$ | 0.24 | 0.34 | -0.10 |
| $a_0 - |\Delta a|$ | 0.12 | 0.17 | -0.05 |
| $a_0$ | 0.00 | 0.00 | 0.00 |
| $a_0 + |\Delta a|$ | -0.12 | -0.17 | 0.05 |
| $a_0 + |\Delta a|$ | -0.24 | -0.34 | 0.10 |